\documentclass[preprint]{elsarticle}

\RequirePackage{amsfonts}
\RequirePackage{amsfonts}
\RequirePackage{mathrsfs}
\RequirePackage{amssymb}
\RequirePackage{amsmath}
\RequirePackage{amsthm}

\usepackage{graphics} 
\usepackage{amsmath,amsthm,amssymb}
\usepackage{xcolor}
\usepackage{enumerate}
\usepackage{graphicx} 
\usepackage{subfig,url}
  
\usepackage{hyperref}
\usepackage{multirow}
\usepackage{multicol}

\usepackage{mathrsfs}
\usepackage{amsfonts}
\usepackage{mathrsfs}
\usepackage{amssymb}
\usepackage{amsmath}
\usepackage{amsthm}

\usepackage{graphbox,graphicx}

\theoremstyle{definition}

\theoremstyle{remark}

\numberwithin{equation}{section}

\begin{document}

\begin{frontmatter}

\title{Current state of nonlinear-type time-frequency analysis and applications to high-frequency biomedical signals}

\author[4,5]{Hau-Tieng Wu}
\ead{hauwu@math.duke.edu}

\address[4]{Department of Mathematics and Department of Statistical Science, Duke University, Durham, NC, USA}
\address[5]{Mathematics Division, National Center for Theoretical Sciences, Taipei, Taiwan}

\begin{abstract}
Motivated by analyzing complicated time series, nonlinear-type time-frequency analysis became an active research topic in the past decades. Those developed tools have been applied to various problems.
In this article, we review those developed tools and summarize their applications to high-frequency biomedical signals.
\end{abstract}


\end{frontmatter}

\section{Introduction}

Recently, complicated high-frequency (multivariate) oscillatory time series has attracted a lot of research interest. This kind of time series is characterized by being composed of multiple oscillatory components with time-varying features, like frequency, amplitude and oscillatory pattern, and contaminated by various kinds of randomness like noise and artifact. 

\begin{figure*}[bpt!]
	\includegraphics[trim=15 90 50 90,clip,width=0.99\textwidth]{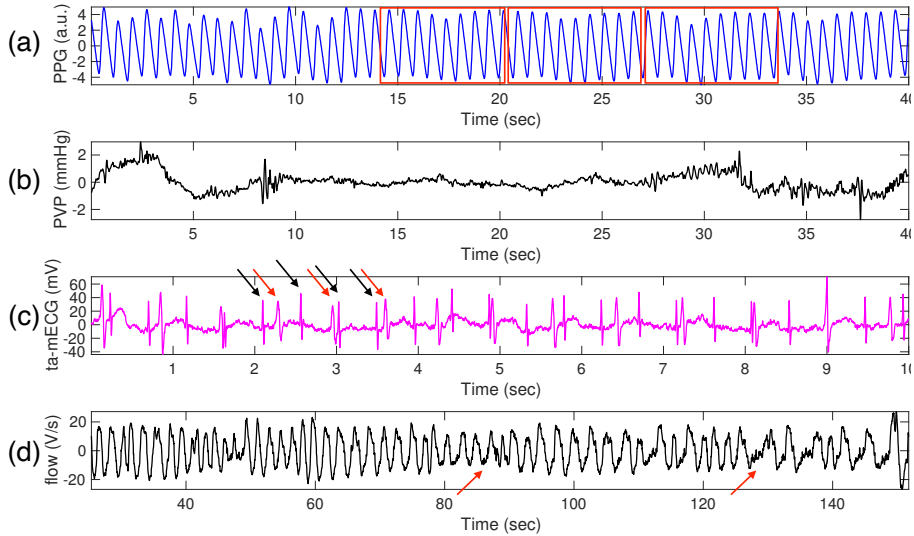}
\caption{Four typical biomedical signals. From top to bottom: the photoplethysmogram (PPG), the peripheral venous pressure (PVP), the trans-abdominal maternal electrocardiogram (ta-mECG), and the respiratory flow signal. In the PPG signal, the red boxes enhance the ``low frequency'' oscillation, which is related to the respiration induced intensity variability. In the ta-mECG, the red arrows indicate the maternal cardiac cycles, and the black arrows indicate the fetal cardiac cycles. In the respiratory flow signal, the red arrows indicate the cardiogenic artifacts, which is a higher frequency oscillation (about 1 second per cycle) compared with the respiratory oscillation (about 5 seconds per cycle).}\label{fig:example}
\end{figure*}

To have a sense of the challenge we encounter, here we present some representative biomedical signals. 
In the photoplethysmography (PPG) signal \cite{shelley2007photoplethysmography}, the signal is composed of a respiratory component and a hemodynamic component. Common goals include calculating oxygen saturation \cite{shelley2007photoplethysmography}, extracting heart and respiratory rate \cite{cicone2017nonlinear} and hence analyzing the heart rate variability (HRV) \cite{lu2008can} and even breathing pattern variability (BPV) \cite{Benchetrit:2000} that represents complicated interaction among various physiological systems \cite{TaskForce:1996}, and even estimating blood pressure \cite{slapnivcar2019blood}. In \ref{fig:example}(a), the red boxes indicate the respiratory cycles.
Another hemodynamic signal is the peripheral venous pressure (PVP) \cite{wardhan2009peripheral}. While it is ubiquitous, it is less studied, probably due to its low signal-to-noise ratio (SNR), and its oscillatory pattern is sensitive to the physiological status. Like PPG, the PVP signal is usually composed of a respiratory component and a hemodynamic component. The goal of analyzing PVP is studying topics related to hypovolemia or in general fluid status \cite{alian2014impact}, or provides surrogate heart or respiratory rate information. In \ref{fig:example}(b), it is not easy to visualize the hemodynamic oscillation due to the low SNR.
The trans-abdominal maternal electrocardiogram (ta-mECG) is composed of one maternal electrocardiogram (ECG) and one fetal ECG \cite{Sameni2010} (multiple fetal ECGs if multiple pregnancy). The ta-mECG is usually contaminated by various artifacts, like motion artifact and uterus contraction, and the fetal ECG is usually of low SNR. One common goal is extracting the fetal ECG \cite{Sameni2010}, and hence study the fetal HRV \cite{lobmaier2019fetal} and fetal arrhythmia \cite{behar2019noninvasive}. In \ref{fig:example}(c), the red and black arrows indicate the maternal and fetal ECG, respectively.
There are various ways to monitor breathing, for example, the  flow signal or the chest/abdominal movements. These signals are usually contaminated by the cardiogenic artifact \cite{seppa2011method}. When the cardiogenic artifact exists, the signal is composed of two components. In clinics, signal analysis interests include quantifying the BPV for patient health evaluation \cite{bien2011comparisons}, recycling the heart rate information \cite{lu2019recycling}, or detecting apnea events during sleep \cite{Poupard2008} or sedation procedure \cite{lo2020hypoventilation}.

From these far-from-exhaustive examples, we could summarize some common interests and challenges when we encounter biomedical signals. 
We would like to decompose the signal into multiple components that represent different physiological systems, which we may call the ``single channel blind source separation (scBSS)'' problem when there is only one channel. We have interest in quantifying the physiological status by further analyzing dynamics of each component; for example, how the signal oscillatory pattern changes, or how fast/strong the signal oscillates over time. Proper quantifications might provide useful physiological dynamics information for clinical decision making. 
However, it is nontrivial toward the above goals, and we have at least the following challenges. 
First, each oscillatory component might have time-varying amplitude and frequency, and even the non-sinusoidal oscillatory pattern changes from time to time. This fact limits traditional signal processing tools. 
Second, the signal might have low SNR, and contaminated by various kinds of unexpected artifacts. Even worse, the oscillatory components may not always exist; that is, it might exist for a while, disappear for a while, and then appear again. 
Third, in many problems, we only have one channel, which limits the tools we can apply. 
Fourth, when the signal is ultra-long, for example, an ECG signal continuously recorded for 14 days, how to numerically efficiently extract useful information, or even visualize the signal.

We mention that in the statistics society, handling this kind of challenging time series is broadly known as the {\em seasonality} problem \cite{Brockwell_Davis:2002}. Many time series tools have been developed, but most of them are {\em not} aiming directly for the above-mentioned challenges. These tools can be roughly categorized into three types: the time domain approach \cite{Oh_Nychka_Brown_Charbonneau:2004,Genton_Hall:2007,Bickel_Kleijn_Rice:2008,Park_Ahn_Hendry_Jang:2011,de2011forecasting}, the frequency domain approach \cite{Fisher1929,Hannan1961,Chiu1989,hall2006using}, or the evolving spectra approach \cite{priestley1965evolutionary,priestley1996wavelets,Dahlhaus:1997,Adak1998,Nason2000,zhou2013heteroscedasticity,zhou2014inference}. This is a far-from-exhaustive list, and we refer readers with interest to those literature and citations therein. 
In the signal processing society, in the past decades, several time-frequency (TF) analysis tools have been proposed to handle these challenges, and tackling seasonality is a special case.
From the data analysis perspective, TF analysis is a set of tools aiming to extracting useful information, or design statistics, from the time series. The outputs are for upcoming statistical analysis, like change point detection, forecasting, etc. While TF analysis and time series analysis are so closely-related, however, there are not much interaction up to recently, except those methods considered in the evolving spectra approach \cite{priestley1965evolutionary,priestley1996wavelets,Dahlhaus:1997,Adak1998,Nason2000,zhou2013heteroscedasticity,zhou2014inference}. In these work, the linear-type or bilinear-type TF analysis tools or similar ideas are used. The lack of combining TF analysis and statistical analysis opens a door for many research opportunities.

Since TF analysis is a huge field, in this review, based on the authors' limited knowledge, we exclusively review some selected models and {\em nonlinear-type} TF analysis tools and mention their application in high-frequency biomedical signals.

\section{Model}

The most common model to describe an oscillatory signal is the {\em harmonic model}; that is, the signal is composed of finite sinusoidal functions with fixed amplitude and frequency: $Y(t)=\sum_{l=1}^LA_l\cos(2\pi\xi_l t+\varphi_l)+\Phi(t)$, where $L\in \mathbb{N}$ is the number of oscillatory components, $t\in\mathbb{R}$ is time, $A_l>0$, $\xi_l>0$ and $\varphi_l\in \mathbb{R}$ are the {\em amplitude}, {\em frequency} and {\em global phase} of the $l$-th oscillatory component respectively, and $\Phi(t)$ is a zero-mean random process. It is also possible to consider a purely stochastic model, like the seasonal autoregressive model \cite{Brockwell_Davis:2002}. This direction has a lot of extensions, and we will not focus on it.

There are two common generalizations of the harmonic model. The first one is the {\em analytic model} widely applied in many engineering problems \cite{Gabor:1946,vanderPol:1946,Picinbono:1997}. A signal satisfies the analytic model if its Fourier transform is supported on the positive frequency axis. The second one is the {\em adaptive harmonic model} (AHM) considered in \cite{DaLuWu2011}, where the amplitude of the $l$-th component is replaced by a positive function $A_l(t)$ called the {\em amplitude modulation} (AM), and the frequency and global phase are replaced by a monotonically increasing function $\phi_l(t)$ called the {\em phase function}. The derivative of $\phi_l(t)$ is thus positive, describing how fast the signal oscillates at time $t$, which we call the {\em instantaneous frequency} (IF). 
We call $A_l(t)\cos(\phi_l(t))$ an {\em intrinsic mode type} (IMT) function, which is motivated by modeling the {\em intrinsic mode function} (IMF) considered in \cite{Huang_Shen_Long_Wu_Shih_Zheng_Yen_Tung_Liu:1998} as the output of the {\em empirical mode decomposition} (EMD). In \cite{Hou_Shi:2011}, a mathematical definition of IMF is given, which is related but different from the IMT function. The identifiability of the AHM function has been extensively discussed in \cite{Chen_Cheng_Wu:2014}.

In practice, the oscillation is usually non-sinusoidal. Motivated by this fact, the notion of {\em wave-shape} function is proposed in \cite{Wu:2013}. A wave-shape function is an $1$-periodic function. With the wave-shape function, the AHM is generalized to the {\em adaptive non-harmonic model} (ANHM). A function satisfies the ANHM if it can be represented as $Y(t)=\sum_{l=1}^KA_l(t)s_l(\phi_l(t))+\Phi(t)$, where $s_l$ is a wave-shape function. 
In \cite{lin2016waveshape,XuYangDaubechies:2018}, this ANHM is further generalized to capture the time-varying wave-shape and dynamics.

With these models, the signal processing missions mentioned in Introduction could thus be formulated as estimating the time-varying amplitude, frequency or oscillatory pattern, and statistical properties of the random process, and quantifying their dynamics, under proper assumptions of the chosen model. For different analysis purposes, the model might be slightly changed. For example, if our interest is change point detection, the model should be adapted to describe change points. Due to page limitation, we will not exhaustively mention all possibilities.

Before closing this section, we mention that in 2019, inspired by analyzing pathophysiological signals, the manifold model was considered to further generalize the ANHM. Specifically, we view each oscillation as a point on a manifold, which is called the {\em wave-shape manifold} \cite{Lin_Malik_Wu:2019}. The recorded signal is then viewed as a realization of the trajectory on the manifold. With the help of diffusion geometry based algorithm like diffusion maps \cite{coifman2006diffusion}, the dynamics hidden in the signal can be quantified for further analysis. While this approach can be incorporated into the TF analysis framework via the spectral geometry, the relationship is out of the scope of this review and we will not further explore, except mentioning that it has been applied to remove stimulation artifact from the direct cortical stimulation \cite{Alagapan_Shin_Frohlich_Wu:2018}, the f-wave extraction for patients with atrial fibrillation \cite{malik2017single}, the fetal ECG extraction from the ta-mECG for arrhythmia screening
\cite{su2019recovery}, the ultra-long physiological time series visualization \cite{Wang_Wu_Huang_Chang_Ting_Lin:2019}, etc.

\section{Current state of nonlinear-type time-frequency analysis}

Broadly speaking, TF analysis is a branch of signal processing aiming to analyze signal from both time and frequency perspectives simultaneously. A general idea underlying TF analysis is studying the ``oscillatory behavior'' ``locally'' at each time point. 
%
%
It converts a given signal, which is a one dimensional function, to a {\em TF representation} (TFR), which is a two dimensional function. With the TFR, we can carry out various missions, like extracting oscillatory information, decomposing the signal, etc. Different TF analysis tools lead to different TFR's, and the notion of ``good'' TFR depends on application. 
The notion of ``local'' also varies from one to another; usually it is quantified by how we truncate a signal into pieces, but it can also be quantified by other ways; for example, the roots distribution of the associated analytic function.

TF analysis techniques can be roughly classified into three types -- linear, bilinear and nonlinear. Examples of linear-type TF analyses include short-time Fourier transform (STFT) and continuous wavelet transform (CWT), and examples of bilinear-type TF analyses include Wigner-Ville distribution (WVD) and more generally the Cohen's class or affine class. 
While linear-type and bilinear-type TF analyses have been extensively studied, they suffer from various limitations. For example, the uncertainty principle is inevitable in linear-type TF analyses, which leads to blurring or smearing out of their TFR's \cite{Ricaud2014,Flandrin:1999}. Also, the results usually depend on the chosen ``atoms'' \cite{Flandrin:1999} and hence not adaptive to the intrinsic structure of the signal. 
The application of WVD is limited by the interference and positivity issues \cite{Flandrin:1999}. While the Cohen's class is designed to resolve these limitations via a kernel smoothing on the TFR, the kernel smooth step might generate unwanted artifacts, the resulting TFR is not adaptive to the signal, and usually decomposing the signal is difficult. We refer readers with interest in linear-type or bilinear-type TF analyses to \cite{Flandrin:1999}.

Nonlinear-type TF analysis aims to resolve these issues encountered in linear-type and bilinear-type TF analyses and depict the signal in a more data-driven way. In a nutshell, it takes more information from the signal to modify the linear-type or bilinear-type TF analyses, so that various signal processing missions can be further carried out. Below, we review various techniques proposed in the past decades chronologically.

\subsection{Reassignment method}

This idea can be traced back to Kodera, et. al. \cite{Kodera1976,Kodera_Gendrin_Villedary:1978} in 1976, where the phase and group delay information of the STFT was applied to sharpen the TFR by reallocating the spectrogram to the right place. In 1995, the idea was further extended to bilinear-type TF or time-scale distributions in \cite{Auger_Flandrin:1995, Chassande-Mottin_Auger_Flandrin:2003}, which is now referred to as the {\em reassignment method} (RM). The basic idea is 
reallocating the TFR to the center of gravity of these energy contributions associated with each TF point.
It is known that the local maxima, or the gravitational center, of the TFR is directly related to the IF \cite{Delprat_Escudie_Guillemain_Kronland-Martinet_Tchamitchian_Torresani:1992}. Thus, the TFR is sharpened for further analysis. 
In \cite{flandrin2013note}, the RM behavior on the Hermite functions was reported; in \cite{chassande1998statistics}, the statistics of the RM was studied. 
Few variations were proposed in the coming few years, including the differential reassignment \cite{Chassande-Mottin_Daubechies_Auger_Flandrin:1997} and adjustable reassignment \cite{Auger_Chassande-Mottin_Flandrin:2012}. To stabilize the noise impact, in 2007 \cite{Xiao_Flandrin:2007}, the RM is combined with the multitaper (MT) technique \cite{Thomson:1982,Babadi_Brown:2014}. The main focus of RM is sharpening the TFR, so decomposing a signal is still challenging. This tool has been applied to study cardio-pulmonary coupling (CPC) \cite{Orini_Bailon_Mainardi_Laguna_Flandrin:2012} and autonomic activity dynamics during general anesthesia \cite{lin2011analyzing,Lin:2015Thesis}. 
See \cite{Auger_Flandrin_Lin_McLaughlin_Meignen_Oberlin_Wu:2013} for a review of this topic.

\subsection{Empirical mode decomposition}

In 1996, a novel signal processing idea, called EMD \cite{Huang_Shen_Long_Wu_Shih_Zheng_Yen_Tung_Liu:1998}, was proposed aiming to obtain a better TFR. The basic idea of EMD is the {\em sifting process}. We first find the upper and lower envelops via applying the spline interpolation over local maxima and local minima respectively, and remove the mean of the upper and lower envelops from the signal. By repeating the same process on the resulting signal until stabilization, we obtain the first IMF, which is the fastest oscillating component, and the slowly oscillating residue.
Then, iteratively apply the sifting processing on the slowly oscillating residue until the termination criteria are met. We thus decompose the signal into several IMFs. 
With those IMFs, the authors proposed to apply Hilbert transform, or other approaches like the quadrature method \cite{huang2009instantaneous}, to estimate the AM and IF of each IMF, and generate the TFR called {\em Hilbert spectra}. After its appearance, EMD has been widely applied to various practical problems and inspired many research interests -- according to Google Scholar, the original publication have been cited more than 20,000 times up to March, 2020. 
A lot of variations have been proposed, like the ensemble empirical mode decomposition (EEMD) \cite{wu2009ensemble}. 
Although it has been widely applied, however, there have been various evidence indicating that EMD might generate misleading results. Moreover, except some rigorous explorations like \cite{flandrin2004empirical,rilling2007one,vatchev2008decomposition}, its mathematical foundation is still lacking, particularly a satisfactory analysis of the sifting process. Due to these limitations, the user should be careful when applying EMD for scientific researches and interpretations, and we will not further review this method, except the above far from complete citations.  

\subsection{Blaschke decomposition}

In 2000, the Blaschke product \cite{Garnett:1981}, a complex analysis result, was applied to analyze signals \cite{Nahon:2000Thesis} under the analytic function model \cite{Gabor:1946,Picinbono:1997}. In \cite{coifman2017carrier}, the authors extended the algorithm to handle non-analytic function, and nominated this Blaschke product based approach the {\em Blaschke decomposition} (BKD). The first key observation \cite{Nahon:2000Thesis} is that the Fourier series expansion of an analytic function is indeed an iterative decomposition procedure. In the $k$-th iteration, we remove $0$ as a root, which lead to the $k$-th Fourier series. 
The second key observation is that the roots of an analytic function inside the unit disk describe when and how fast the signal oscillates. In BKD, these two observations are combined together to decompose an analytic signal. Specifically, in each iteration, all roots inside the unit disk are removed, and hence the $k$-th component. 
We can then apply any suitable TF analysis to analyze each decomposed component. By superimposing all TFRs of all decomposed components, we obtain the TFR for the signal \cite{coifman2017carrier}. 
There have been several theoretical supports for the BKD under different mathematical models; for example, the convergence in $H^2$ (Hardy space) was given in \cite{qian2010intrinsic}, the convergence in a large family of function spaces, including $H^2$ and all Sobolev spaces, was given in \cite{coifman2017nonlinear}, an extension to $H^p$ (Hardy space) was shown in \cite{coifman2019phase}, and the root behavior was studied in \cite{steinerbergerzeroes}. However, a rigorous proof of the numerically observed exponential convergence property is still lacking. Recently, an iterative approach is applied to enhance the algorithm \cite{lukianchikov2019iterative}. We mention that a similar idea was considered in \cite{qian2010intrinsic} based on the Malmquist-Takenaka system \cite{takenaka1925orthogonal,eisner2014discrete}, where the authors called the algorithm {\em adaptive Fourier decomposition} \cite{qian2011algorithm}. We mention that it is possible to consider other general orthogonal functions \cite{pap2006voice,feichtinger2013hyperbolic} to achieve the decomposition. We will not further survey studies in this direction.

\subsection{(Adaptive locally) iterative filtering}

In 2009, motivated by the novel idea beyond sifting processing, the iterative filtering decomposition (IFD) was proposed \cite{Lin_Wang_Zhou:2009,wang2012iterative} with several theoretical supports \cite{Huang_Wang_Yang:2009,cicone2020study}. The sifting process is captured in the IFD by an iterative high-pass filter. After the iterative high-pass filter, the fast oscillating component is extracted, and the signal is decomposed by repeating this process. With the decomposed components, the TFR can be generated by the same process proposed in EMD, or by applying any suitable TF analysis to each component and then superimpose the TFR of each component In \cite{cicone2016adaptive}, the iterative filtering algorithm is generalized to the {\em adaptive local iterative filtering} (ALIF) algorithm \cite{cicone2016adaptive}. Unlike the iterative filtering algorithm, in ALIF the prior knowledge of instantaneous frequency is taken into account to design a time-varying high-pass filter. Some theoretical analysis of ALIF is available \cite{cicone2019spectral}. The iterative filtering has been applied to automatically classify sleep stages from the EEG signal \cite{sharma2017automatic}.

\subsection{Sparsity approach}

In 2011, inspired by the EMD and the compressed sensing theory, an optimization approach with the sparsity constraint to adaptively quantify the oscillatory behavior was proposed in \cite{Hou_Shi:2011,Hou_Shi:2013a}, which led to the sparse time-frequency representation (STFR). The authors constructed a large dictionary containing all possible oscillatory basis functions, and fit a given signal by finding the sparsest representation from the designed over-complete dictionary. Extensive theoretical results supporting this approach have been established; including the convergence analysis \cite{Hou_Shi:2013a,Hou_Shi_Tavallali:2014} and the relationship between the dictionary and the second order differential equation
\cite{Hou_Shi_Tavallali:2013}.
The STFR approach was extended to study the ``intra-wave frequency modulation'' in
\cite{tavallali2014extraction,HouShi:2016}, which is a model in parallel with the wave-shape model. The convergence property of this extension was shown in \cite{tavallali2015convergence}, and it was applied to the hemodynamic waveform analysis 
\cite{Pahlevan_Tavallali_Rinderknecht_Petrasek_Matthews_Hou_Gharib:2014} and insulin resistance analysis \cite{petrasek2015intrinsic}.

\subsection{Synchrosqueezing transform}

In 2011, motivated by studying the sifting process in EMD, the {\em synchrosqueezing transform (SST)} proposed in 1995 \cite{Maes:1995,DaubechiesMaes1996}, a tool to sharpen the TFR by CWT, was reconsidered and extended to decompose signal \cite{DaLuWu2011} under the AHM. 
The name SST joins ``synchro-'', evoking the causal preservation for the decomposition, with ``squeezing'', suggesting that the chosen TFR is sharpened via reassignment along the frequency axis. The reassignment step depends on the stationary phase approximation \cite{Delprat_Escudie_Guillemain_Kronland-Martinet_Tchamitchian_Torresani:1992} for the CWT. From the perspective of reassignment, the SST can be viewed as a special RM. 
After \cite{DaLuWu2011}, the SST was extended to squeeze the STFT \cite{Wu:2011Thesis}, the S-transform \cite{Huang_Zhang_Zhao_Sun:2015}, the wave package transform \cite{Yang:2014}, the chirplet transform \cite{zhu2019multiple}, and various other extensions \cite{berrian2017adaptive}. 
Specifically, the second \cite{Oberlin_Meignen_Perrier:2015} or higher order \cite{pham2017high} phase information is taken into consideration to further sharpen the TFR, particularly when there are non-trivial chirp components. In \cite{ahrabian2015synchrosqueezing,MERT2018106}, the SST is generalized to handle multi-dimensional time series.
The SST has been widely applied to analyze biomedical signals. For example, estimating instantaneous heart rate from non-contact PPG \cite{Wu_Lewis_Davila_Daubechies_Porges:2015}, extracting instantaneous respiratory rate from PPG \cite{dehkordi2018extracting,jarchi2018validation}, diagnosis of paroxysmal atrioventricular block from single-lead ECG \cite{kabir2014development}, automatic QRS complex detection from single-lead ECG \cite{doi:10.1080/03772063.2016.1221744,Herry_Frasch_Seely_Wu:2017}, quantification of respiratory and cardiac synchronization \cite{iatsenko2013evolution,hemakom2017quantifying}, emotional state prediction from EEG \cite{ozel2019synchrosqueezing,MERT2018106}, spindle or K-complex detection during sleep \cite{kabir2015enhanced,ghanbari2017k}, automatic sleep stage annotation from EEG \cite{Wu_Talmon_Lo:2015}, EEG signal sonification \cite{rutkowski2014multichannel}, the extraction of ECG-derived respiration signal for patients with atrial fibrillation \cite{Wu_Chan_Lin_Yeh:2014}, automatic sleep apnea annotation from respiratory efforts \cite{Lin_Wu_Hsu_Wang_Huang_Huang_Lo:2016}, BPV quantification \cite{Baudin_Wu_Bordessoule_Beck_Jouvet_Frasch_Emeriaud:2014} and its application to mechanical ventilation weaning prediction \cite{Wu_Hseu_Bien_Kou_Daubechies:2013}, HRV quantification for differential effects of sevoflurane anaesthesia \cite{Lin_Wu_Tsao_Yien_Hseu:2014} bronchoscopic sedation control \cite{lo2020hypoventilation}, heart and respiration signals from doppler radar sensing \cite{yavari2016synchrosqueezing,zhao2017noncontact}, etc.
Recently, for the sake of applying the nonlinear-type TF analysis for statistical inference, the asymptotical distribution of SST under null and non-null setup is studied in \cite{SoWuZh2019}. This is probably one of few results in this direction up to date, in addition to some earlier contribution in \cite{chassande1998statistics}.

\subsection{Scattering transform}
In 2012, motivated by studying the convolutional neural network (CNN), the {\em scattering transform} (ST) was proposed \cite{Mallat:2012}. Following the multi-layer structure of CNN, the ST computes the wavelet transform and its modulation of a given signal, and iteratively applies the same process to the modulation to build up a cascade of coefficients as features for the signal. The main feature of the ST is its robustness to local deformation, which is commonly encountered in physiological signals. Its theoretical property on intermittent process analysis has been reported in \cite{bruna2015intermittent}, and it has been applied to study the spectral property of a given signal, called deep scattering spectrum \cite{anden2014deep}. Recently, a generalization called {\em time-frequency scattering} is proposed in \cite{anden2019joint}. It has been applied to study sleep dynamics \cite{liu2020diffuse}, fetal HRV \cite{chudavcek2013scattering}, etc.

\subsection{Concentration of frequency and time -- generalized multitaper}

While the SST is a nonlinear method, its robustness to noise is first shown in \cite{Brevdo_Fuckar_Thakur_Wu:2012} when the SST is applied to CWT, and later extended to different types of noise, including non-stationary and heteroscedastic noises \cite{Chen_Cheng_Wu:2014}. See  \cite{yang2018statistical} for a noise analysis of synchrosqueezed wave packet transform. However, when the SNR is low, the SST may still fail. In 2016, viewing the fact that the traditional MT considered in \cite{Xiao_Flandrin:2007} is limited by the Nyquist restriction in the TF domain \cite{Daubechies:1988}, a generalized MT is proposed \cite{DaWaWu2016}. In a nutshell, when two windows are not orthogonal, the nonlinearity of the chosen nonlinear-type TF analysis, like SST, helps ``de-correlate'' the noise structure, and hence achieves a stabilization of the noise impact. The algorithm is coined {\em concentration of frequency and time} (ConceFT). 
ConceFT has been applied to various clinical problems; for example, the autonomic reaction to noxious stimulation \cite{Lin_Wu:2016}, the dynamics of the transient emission otoacoustic signal \cite{Wu_Liu:2018}, and HRV and QT interval variability quantification for ultra-long ECG signal \cite{Wu_Soliman:2018}.

\subsection{de-shape}

In 2017, in order to handle the non-sinusoidal oscillatory signals,
the {\em de-shape} algorithm was introduced under the ANHM \cite{lin2016waveshape}.
The key observation is that the wave-shape function can be represented as a summation of sinusoidal functions with integer frequencies. Thus, in the frequency domain, the power spectrum is oscillatory, with the period the same as the frequency of the original signal. The resulting TFR, for example the spectrogram, is thus complicated and difficult to interpret, particularly when there are multiple non-sinusoidal oscillatory components.
This observation rings the bell of {\em cepstrum} \cite{oppenheim2004frequency}, which is the spectrum of the nature log of the power spectrum of the signal. 
Based on the reciprocal relationship between frequency and period, the cepstrum contains the period information of each oscillatory components. 
The ``frequency'' axis of cepstrum is called {\em quefrency} \cite{oppenheim2004frequency}, whose unit is time. 
The de-shape algorithm is composed of two steps. First, the cepstrum is extended to the windowed version, which is called the {\em short-time cepstral transform} (STCT). Second, the STCT is converted by reversing the quefrency axis, which is called the {\em inverse STCT} (iSTCT). The iSTCT contains the frequency information and its division, due to the reciprocal relationship between frequency and period. Since the common information in the STFT and iSTCT is the fundamental IF, the iSTCT is taken as a nonlinear mask to remove multiples from the STFT, and hence the final TFR. If needed, the phase information can be further applied to sharpen the TFR, and hence the de-shape SST. The resulting TFR thus contains only the fundamental IF's of all oscillatory components.
The de-shape STFT has been applied to several biomedical problems. For example, fetal ECG extraction from the ta-mECG
\cite{Su_Wu:2016b,LiFraschWu2017} and hence maternal stress evaluation \cite{lobmaier2019fetal} and fetal arrhythmia screening \cite{su2019recovery}, instantaneous heart and respiratory rates estimation from the PPG signal \cite{cicone2017nonlinear}, cardiogenic artifact recycle from respiratory signal \cite{lu2019recycling}, analysis of sawtooth artifact in patient monitor \cite{lin2019unexpected}, etc.
While applying the de-shape algorithm, several limitations were found, which motivated replacing the iSTCT \cite{lin2016waveshape} by the periodic transform \cite{sethares1999,PPV3}, which leads to a novel TF analysis tool called {\em Ramanujan de-shape} (RDS). The robustness analysis of RDS and preliminary ta-mECG analysis is provided in \cite{RDS2000}.

\subsection{Other approaches}

There are many other nonlinear approaches that can be categorized as the nonlinear-type TF analysis but not summarized above due to the page limit. A non-exhaustive list includes the adaptive multiwindow and multilayered Gabor expansion \cite{Jaillet_Torresani:2007}, partial differential equation based approach \cite{wang2012mode}, the non-local spectrogram approach for the denoise purpose \cite{Galiano_Velasco:2014}, the TF filtering based on spectrogram zeros \cite{Flandrin:2015}, the nonlinear mode decomposition \cite{IMCStefanovska:2015}, the approximation approach to directly estimate IF and AM for signal decomposition \cite{chui2016signal,chui2016data}, the entropy-based window selection to further sharpen the TFR \cite{sheu2017entropy}, the convex optimization approach to construct the TFR by fitting the signal, called Time-frequency bY COnvex OptimizatioN (Tycoon) \cite{kowalski2018convex}, Joint Estimation of Frequency, Amplitude and Spectrum (JEFAS) for time warping and amplitude modulation signals \cite{meynard2018spectral}, etc.

\section{Discussion and Conclusion}\label{sec: conclusion}

We provide a review of current state of nonlinear-type TF analysis and focus on summarizing its application to high-frequency biomedical signals. This is a broad field under active development, and the developed TF analysis tools have been applied to many different fields. We hope this review could serve as a starting point of entering this field. 


\bibliographystyle{siam} 
\bibliography{conceft}

\end{document}